\documentclass[showpacs,amsmath,amssymb,onecolumn,prx,reprint,superscriptaddress,floatfix]{revtex4-2}

\usepackage[dvips]{graphicx}
\usepackage{subfigure}
\usepackage{amsfonts}
\usepackage{amssymb}
\usepackage{amscd}
\usepackage{amsmath}
\usepackage{enumerate}
\usepackage{epsfig}
\usepackage{subfigure}
\usepackage{bm}
\usepackage{xcolor}
\usepackage{amsthm}
\usepackage{framed}
\usepackage{multirow}
\usepackage{mathrsfs,amssymb}
\usepackage{mathtools}
\usepackage{physics}
\usepackage{upgreek}
\usepackage[most]{tcolorbox}

\usepackage[version=4]{mhchem}
\usepackage{chemformula}

\begin{document}
\title{Real-time vacuum-state quantum random number generator on a chip}

\begin{abstract}
  Quantum random number generators (QRNGs) produce true random numbers, which are guaranteed by the fundamental principles of quantum physics.
  Miniaturization of QRNGs is crucial for a wide range of communication and cryptography applications.
  Here, we first report a fully functional QRNG chip based on vacuum-state fluctuations, with dimensions of 16.6 mm × 7.8 mm.
  The quantum entropy source, which is achieved via hybrid photonic integration with a \ce{SiO2} waveguide, generates raw quantum random numbers.
  The hybrid photonic and electrical components are assembled into a compact ceramic package using system-in-package technology.
  A microcontroller unit acquires the raw data and outputs the processed quantum random numbers via a serial peripheral interface.
  According to the characterization results, the QRNG chip achieves a constant real-time output rate of 5.2 Mbps across the industrial temperature range of -40°C to 85°C, making it suitable for practical applications.

\end{abstract}

\author{Guan-Ru Qiao}
\affiliation{Hefei National Research Center for Physical Sciences at the Microscale and School
of Physical Sciences, University of Science and Technology of China, Hefei 230026, China.}
\affiliation{CAS Center for Excellence in Quantum Information and Quantum Physics,
University of Science and Technology of China, Hefei 230026, China.}

\author{Bing Bai}
\email{baib@ustc.edu.cn}
\affiliation{Hefei National Research Center for Physical Sciences at the Microscale and School
of Physical Sciences, University of Science and Technology of China, Hefei 230026, China.}
\affiliation{CAS Center for Excellence in Quantum Information and Quantum Physics,
University of Science and Technology of China, Hefei 230026, China.}

\author{Zi-Xuan Weng}
\affiliation{Hefei National Research Center for Physical Sciences at the Microscale and School
of Physical Sciences, University of Science and Technology of China, Hefei 230026, China.}
\affiliation{CAS Center for Excellence in Quantum Information and Quantum Physics,
University of Science and Technology of China, Hefei 230026, China.}

\author{Han-Shen Chen}
\affiliation{Hefei National Research Center for Physical Sciences at the Microscale and School
of Physical Sciences, University of Science and Technology of China, Hefei 230026, China.}
\affiliation{CAS Center for Excellence in Quantum Information and Quantum Physics,
University of Science and Technology of China, Hefei 230026, China.}

\author{Wei Zheng}
\affiliation{Henan Shijia Photons Technology Co. Ltd, Hebi 458030, China.}

\author{Zhi-Yuan Zheng}
\affiliation{Henan Shijia Photons Technology Co. Ltd, Hebi 458030, China.}

\author{You-Qi Nie}
\affiliation{Hefei National Research Center for Physical Sciences at the Microscale and School
of Physical Sciences, University of Science and Technology of China, Hefei 230026, China.}
\affiliation{CAS Center for Excellence in Quantum Information and Quantum Physics,
University of Science and Technology of China, Hefei 230026, China.}
\affiliation{Hefei National Laboratory, University of Science and Technology of China,
Hefei 230088, China.}

\author{Jun Zhang}
\email{zhangjun@ustc.edu.cn}
\affiliation{Hefei National Research Center for Physical Sciences at the Microscale and School
of Physical Sciences, University of Science and Technology of China, Hefei 230026, China.}
\affiliation{CAS Center for Excellence in Quantum Information and Quantum Physics,
University of Science and Technology of China, Hefei 230026, China.}
\affiliation{Hefei National Laboratory, University of Science and Technology of China,
Hefei 230088, China.}

\author{Jian-Wei Pan}
\affiliation{Hefei National Research Center for Physical Sciences at the Microscale and School
of Physical Sciences, University of Science and Technology of China, Hefei 230026, China.}
\affiliation{CAS Center for Excellence in Quantum Information and Quantum Physics,
University of Science and Technology of China, Hefei 230026, China.}
\affiliation{Hefei National Laboratory, University of Science and Technology of China,
Hefei 230088, China.}

\maketitle

\section{Introduction}

  Random numbers are widely used in applications such as communication, cryptography, simulation, lottery drawing, and scientific computation.
  Quantum random number generators (QRNGs) produce true random numbers characterized by unpredictability, irreproducibility, and unbiasedness, which are guaranteed by the fundamental principles of quantum physics. 
  Over the last two decades, several QRNG schemes have been demonstrated \cite{herrero2017quantum, ma2016quantum}, including the beam splitter scheme by measuring the path selection of single photons, the time measurement scheme by digitizing the arrival time of single photons \cite{dynes2008high, nie2014practical}, the quantum phase fluctuation scheme by measuring phase fluctuations due to the spontaneous emission of a laser \cite{xu2012ultrafast, nie2015generation, zhou2015randomness, zhang2016note, lei20208, marangon2024fast}, and the vacuum-state scheme by measuring quantum noise fluctuations \cite{gabriel2010generator,shi2016random}.

  In the vacuum-state scheme, quantum noise fluctuations are measured by a homodyne detector.
  Randomness arises from quadrature measurement of the vacuum state, with the uncertainty principle ensuring that the measurement outcomes are inherently unpredictable.
  Recently, several research groups have demonstrated experimental implementations of vacuum-state QRNGs based on photonic integration technology \cite{bai202118, bruynsteen2023100, tanizawa2024real, wang2024compact}.
  To facilitate widespread adoption in communication and cryptography applications, achieving an integrated and miniaturized standalone QRNG is critical.
   
  Vacuum-state QRNGs comprise a quantum entropy source, data acquisition, and a postprocessing component responsible for extracting the final random numbers.
  Several optical waveguide platforms are suitable for integrating the quantum entropy source \cite{wang2020integrated,luo2023recent}, including silicon \cite{siew2021review}, silicon dioxide (\ce{SiO2}) \cite{kawachi1990silica,politi2008silica}, indium phosphide \cite{abellan2016quantum}, gallium arsenide \cite{chang2018heterogeneously}, silicon nitride \cite{roeloffzen2018low}, and lithium niobate \cite{boes2023lithium}. 
  Owing to its advantages of low cost, minimal optical loss, high chemical stability, and ease of fabrication and integration, \ce{SiO2} is widely used for optical waveguides in industrial manufacturing.
  Hybrid photonic integration \cite{kaur2021hybrid,baek2025advanced} is employed to combine waveguides with a light source and detectors to implement the entropy source.
  Furthermore, the entropy source and its associated electronic components are integrated to achieve a chip-level QRNG, for which system-in-package (SiP) technology is applied.
  
  \begin{figure*}[htp]
    \centerline{\includegraphics[width=17.5 cm]{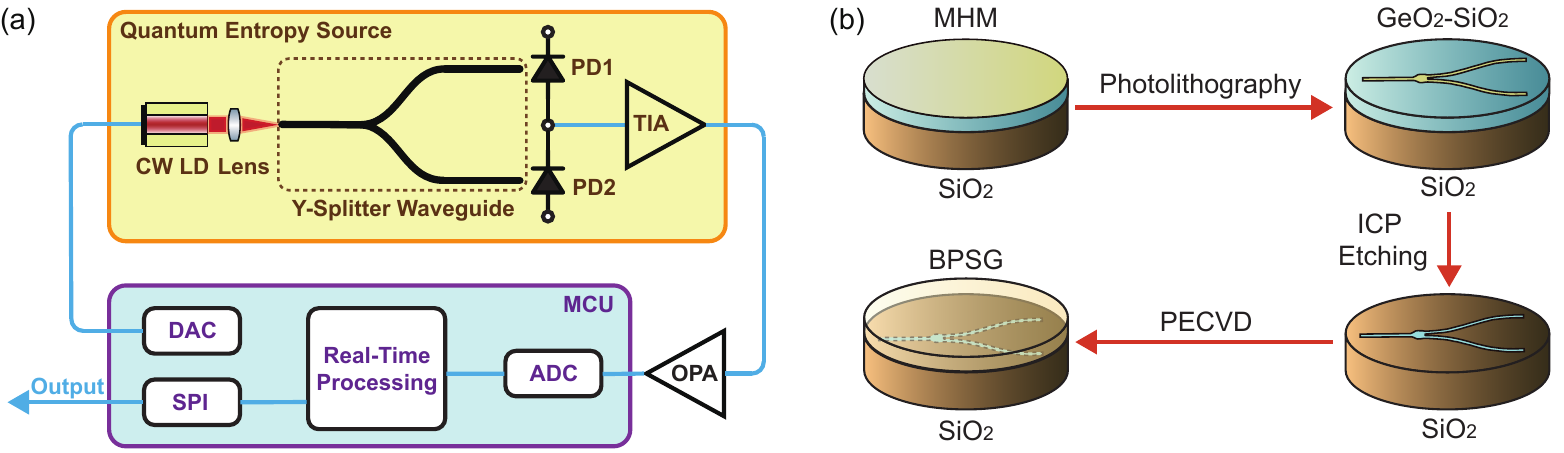}}
    \caption{(a) Schematic diagram of the QRNG chip. (b) Technological process of the \ce{SiO2} Y-splitter waveguide. CW LD: continuous-wave laser diode, PD: photodetector, TIA: transimpedance amplifier, OPA: operational amplifier, MCU: microcontroller unit, ADC: analog-to-digital converter, DAC: digital-to-analog converter, SPI: serial peripheral interface, MHM: metal hard mask, ICP: inductively coupled plasma, PECVD: plasma-enhanced chemical vapor deposition, BPSG: boro phospho silicate glass.}
    \label{figure1}
  \end{figure*} 

  In this work, we present, for the first time, a QRNG chip based on vacuum-state fluctuations, with a real-time output rate of 5.2 Mbps and a size of 16.6 mm × 7.8 mm.
  The vacuum-state entropy source is realized through hybrid photonic integration via a \ce{SiO2} Y-splitter waveguide.
  Both the optical and electronic components of the QRNG chip are integrated into a ceramic package using SiP technology.
  Data acquisition, randomness extraction, and system monitoring are managed by a microcontroller unit (MCU).
  The QRNG chip realizes a stable final random number output rate via a serial peripheral interface (SPI) and passes standard randomness tests across the industrial temperature range of -40°C to 85°C.
  
  \section{Implementation of the QRNG Chip}

  Figure~\ref{figure1}(a) presents a schematic diagram of the QRNG chip based on vacuum-state fluctuations, which encompasses both the quantum entropy source and the electronic components.
  The quantum entropy source is a hybrid photonic component consisting of a Y-splitter waveguide, a continuous-wave laser diode (LD), a lens, photodiodes (PDs), and a transimpedance amplifier (TIA).
  The LD is coupled to the waveguide through a lens.
  The TIA amplifies the photocurrent difference detected at the common electrode of the two PDs.
  The quantum noise signal is further amplified by an operational amplifier (OPA) before being fed into the analog-to-digital converter (ADC) of the MCU.
  A digital-to-analog converter (DAC) controls the LD light intensity via a triode.
  The MCU performs raw data postprocessing and outputs the final quantum random numbers via the SPI peripheral.

  In the vacuum-state QRNG scheme, the randomness arises from quadrature measurement of the vacuum-state fluctuations, with the uncertainty principle guaranteeing the unpredictability of the measurement outcomes.
  In the standard vacuum-state scheme, local oscillator (LO) light is injected into one port of a 50:50 beam splitter while the other port is blocked.
  
  In Y-splitter-based vacuum-state QRNGs, the coherent state is sent into one input port of the Y-splitter, while vacuum states are sent into the other two ports, which are coupled to the homodyne detector.
  The vacuum states sent to the ports of the Y-splitter introduce random fluctuations in the vacuum field, performing the same function as the blocked port of the beam splitter.
  Ivanova \textit{et al.} \cite{ivanova2017quantum} proved that mathematical expressions demonstrate the formal equivalence between a 50:50 beam splitter and a symmetric Y-splitter.  
  Thus, the use of the Y-splitter enables the same evaluation method as that for a beam splitter to be used for the QRNG system.

  \begin{figure*}[ht!]
    \centerline{\includegraphics[width=17 cm]{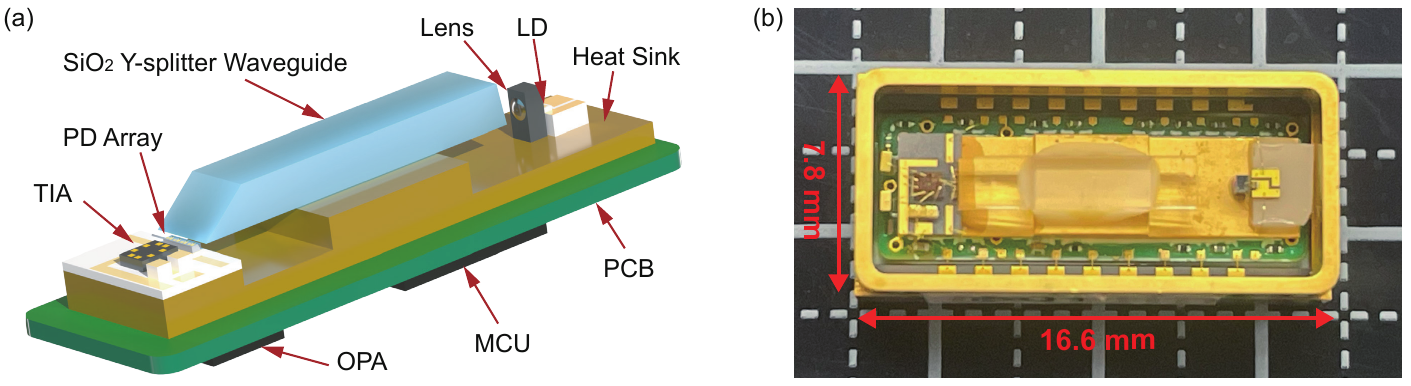}}
    \caption{(a) Interior structure of the QRNG chip. (b) Photograph of the QRNG chip. PCB: printed circuit board.}
    \label{figure2}
  \end{figure*} 
  
  The Y-splitter waveguide in the entropy source is fabricated using \ce{SiO2}-based planar lightwave circuit technology. 
  \ce{SiO2} benefits from well-established and mature manufacturing processes, which ensure high production efficiency and low cost.
  The waveguide fabrication process is illustrated in Fig.~\ref{figure1}(b).
  The \ce{SiO2}-based planar lightwave circuit consists of a structure with three layers: the substrate, core, and upper cladding.
  The \ce{SiO2} layer serves as both the substrate and the lower cladding layer \cite{kawachi1990silica, henderson2009structure}.
  The \ce{GeO2}--\ce{SiO2} core is deposited onto the \ce{SiO2} layer via plasma enhanced chemical vapor deposition (PECVD).
  Germanium-doped \ce{SiO2} has a relatively high refractive index, enabling the fabrication of optical waveguides with a refractive index contrast.

  A metal hard mask is first deposited on the core layer and then coated with a photoresist, which is patterned into a Y-splitter shape.
  After photolithography, the core layer undergoes inductively coupled plasma etching, followed by removal of the metal hard mask.
  Finally, boro-phospho-silicate glass is deposited as the upper cladding via PECVD, completing the waveguide fabrication.
  The refractive indices of the upper and lower cladding layers are both 1.445 at a wavelength of 1550 nm, with a refractive index difference of 0.36\% between the core layer and the cladding layers.
  Through process refinement and iteration, the splitting ratio of the waveguide is optimized to achieve near-perfect symmetry.

  The \ce{SiO2} waveguide is a passive device, therefore, a light source and detectors are required to realize vacuum-state quantum entropy source integration.
  Electronic components are also necessary for data acquisition, postprocessing, and system monitoring in a QRNG system.
  SiP is an advanced integration technology that enables the assembly of diverse components with different functions into a single chip package, facilitating the realization of specific functionalities \cite{khan2023exploring,wang2023review,baek2025advanced}.
  A fully functional chip-scale QRNG, integrating both optical and electronic components, is realized through SiP technology.

  Figure~\ref{figure2}(a) presents the interior structure of the QRNG chip.
  The Y-splitter waveguide, LD, lens, TIA, and PD array are mounted on a Kovar alloy heat sink, which serve as the hybrid photonic entropy source.
  The distributed feedback LD with a center wavelength of 1550 nm is coupled to the waveguide via a lens.
  The LD power is determined from the operating current, with a typical threshold current of 10 mA and a slope efficiency of 0.3 W/A.

  The integrated homodyne detector \cite{raffaelli2018homodyne,bruynsteen2021integrated,ng2024chip,ng2025gigabit} is a key component for practical quantum state measurement in QRNGs.
  The PD array has a responsivity of 0.9 A/W at a wavelength of 1550 nm and features an active area with a diameter of 50 µm, serving as a homodyne detector.
  The high-gain TIA, with a transimpedance gain of 55 k$\Omega$ and low power consumption, is wire bonded to the electrodes of the homodyne detector and amplifies the weak electrical signals from the PDs.

  The heat sink of the hybrid photonic component features stepped height offsets.
  This offset structure allows alignment of the LD, lens, and waveguide on the same horizontal plane, thus ensuring precise coupling.
  The TIA and PD array are mounted on a thin-film circuit board placed on the heat sink to reduce the wire-bonding distance of the TIA.
  The QRNG chip features a stacked configuration, with the entropy source placed on the top layer of a printed circuit board (PCB) and the MCU and OPA on the bottom layer.
  This integrated structure optimizes the spatial utilization within the QRNG chip.

  In the vacuum-state scheme, balanced splitting is essential for generating the quantum noise signal while ensuring an even distribution of the signal across both detection channels. 
  Therefore, calibration of component positions and performance testing are needed during the hybrid photonic component assembly process. 

  An external reference light source is employed during calibration, enabling fine adjustments.
  Current monitoring of the PDs and the power supply to the TIA and LD is carried out through probes. 
  By connecting a current meter to the LD and the PDs, the operating current of the laser and the optical response currents of the PDs are measured.
  The PD array is precisely adjusted to ensure that the active areas of both PDs respond equally to the waveguide emission light.
  The Y-splitter waveguide is affixed to the heat sink after the alignment of the PD array and waveguide is completed.
  When the LD is coupled to the Y-splitter waveguide input port, the PD array is monitored to ensure consistency of the dual-channel response current.
  The position of the lens is adjusted to maximize the PDs’ response current, and the lens is then fixed in place with glue.
  All components are securely fixed onto the heat sink to complete the assembly of the hybrid photonic component.

  In the vacuum-state scheme, imbalances in the homodyne detector can reduce the signal-to-noise ratio between the quantum shot noise and the classical electronic noise.
  Through the precise calibration processes, a 49.2:50.8 splitting ratio is achieved without the need for optical modulation.
  Based on the measured LD operating current and PD photocurrents, the total transmission loss of the photonic chip is approximately 3.2 dB. 
  
  The QRNG chip is finally assembled in a ceramic shell, as shown in Fig.~\ref{figure2}(b). 
  A fully functional vacuum-state chip-scale QRNG is achieved by integrating the entropy source with the electronic components. 
  The PCB with the MCU and OPA is mounted on the step surrounding the ceramic shell.
  After the entropy source is adhered on PCB, wire bonding connects the entropy source, PCB pads, and the ceramic shell pins.
  The DFN-18 package of the ceramic shell with a 1.27 mm pin pitch has dimensions of 16.6 mm × 7.8 mm.
  
  \section{Random data processing}
  In the QRNG chip, the MCU serves as the central unit for random data processing to generate the final quantum random numbers. 
  We choose a microsized MCU with dimensions of 4.2 mm × 3.8 mm for the QRNG chip.
  Built on the Arm Cortex-M7 architecture, the MCU offers advantages such as low cost, ease of development, and energy efficiency.
  The MCU is developed using the C programming language platform and is equipped with diverse system peripherals, including ADCs, DACs, and multiple communication interfaces.
  Considering factors such as chip size, real-time output speed, and cost, the MCU offers an efficient and practical solution.

  \begin{figure}[ht]
    \centerline{\includegraphics[width=8 cm]{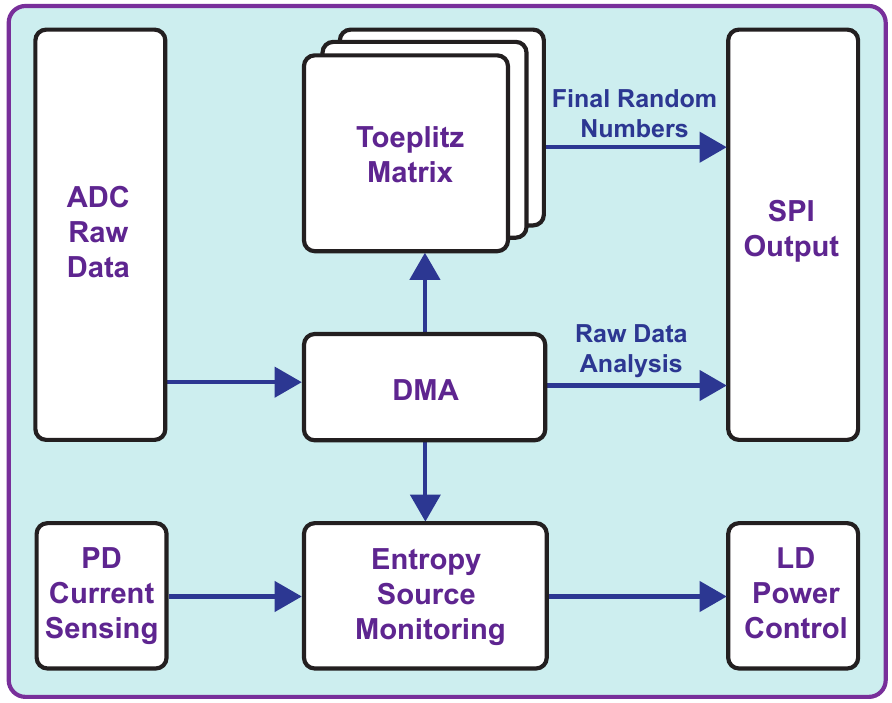}}
    \caption{Functional block diagram inside the MCU. DMA: direct memory access.}
    \label{figure3}
  \end{figure} 

    The functional block diagram of the MCU is shown in Fig.~\ref{figure3}.
    Key QRNG functions, including raw data acquisition, randomness extraction, data output, and entropy source monitoring, are managed by the MCU.
    The original noise signal from the TIA undergoes two-stage amplification via a dual-channel OPA with a 20 MHz gain-bandwidth product. 
    A standard single-ended, noninverting AC amplifier circuit is optimized to match the ADC measurement range.
    Once digitized by the ADC, the raw quantum noise data are transferred to the random access memory of the MCU via the direct memory access (DMA) peripheral.
    Employing DMA for data transfer is a strategic decision aimed at optimizing the system efficiency.
    Given that the MCU postprocessing unit operates in serial computing mode, the use of DMA minimizes MCU involvement during data transmission.
    Offloading this task to the DMA controller significantly reduces system resource consumption, allowing the MCU to allocate more processing power to random data postprocessing.
    The raw data received via DMA can either be sent to the postprocessing unit or output through the SPI peripheral for further analysis.

    The Toeplitz matrix serves as a hashing extractor to refine the raw random data \cite{zhang2016note, ma2013postprocessing}.
    It efficiently transforms raw Gaussian-distributed noise into a uniformly distributed random number sequence.
    The extracted randomness consists of intrinsic randomness from quantum shot noise and randomness from classical noise \cite{zhou2018randomness, yuan2019quantum}. 
    For a Toeplitz matrix of size $m\times n$, $n$ final random bits are extracted by multiplying the matrix by $m$ raw bits. 
    The values of $m$ and $n$ are critical for optimizing the computational efficiency, especially considering the hardware constraints of the MCU. 
    To streamline the processing, both $m$ and $n$ should be integer multiples of 32. 
    This aligns with the MCU architecture, which employs 32-bit registers, thus enabling efficient matrix multiplication operations. 
    The ADC resolution is set to 8 bits, which facilitates the combination of raw data into 32-bit inputs for the Toeplitz matrix computation unit. 

    To further enhance the randomness quality, multiple Toeplitz matrices are initialized in the MCU flash memory.
    These matrices are rotated during postprocessing, meaning that different matrices are used for successive batches of raw data. 
    This rotation prevents patterns or correlations that might arise from the repeated use of a single matrix.
    A portion of the postprocessed random number data is used to update the seeds of the rotation matrices, thereby improving the security of the extractor.
    The final quantum random number data are output via the standard SPI protocol, ensuring backward compatibility with various backend reception rate requirements. 
    
    The MCU incorporates an entropy source monitoring function to ensure the continuous and stable operation of the QRNG chip.
    During QRNG chip initialization, the DAC of MCU scans the LD voltage through the triode.
    The operating point is determined during this calibration process based on the distribution of the raw data.
    The PD currents are collected and serve as a reference for adjusting the laser intensity.
    The MCU evaluates PD currents and the raw data min-entropy to realize the entropy source health monitoring function.   
    
    The LD operating point is recalibrated through feedback adjustments based on the randomness threshold and PD currents.
    The adaptive adjustment algorithm of the MCU improves the chip stability and environmental adaptability.
    This MCU-based system monitoring ensures a constant real-time output rate over extended periods within the industrial temperature range.
    The entropy source monitoring parameters and threshold settings are determined based on the results of the QRNG chip characterization.
  
  \section{Characterization of the QRNG chip}

  \begin{figure*}[ht]
    \centerline{\includegraphics[width=16.5 cm]{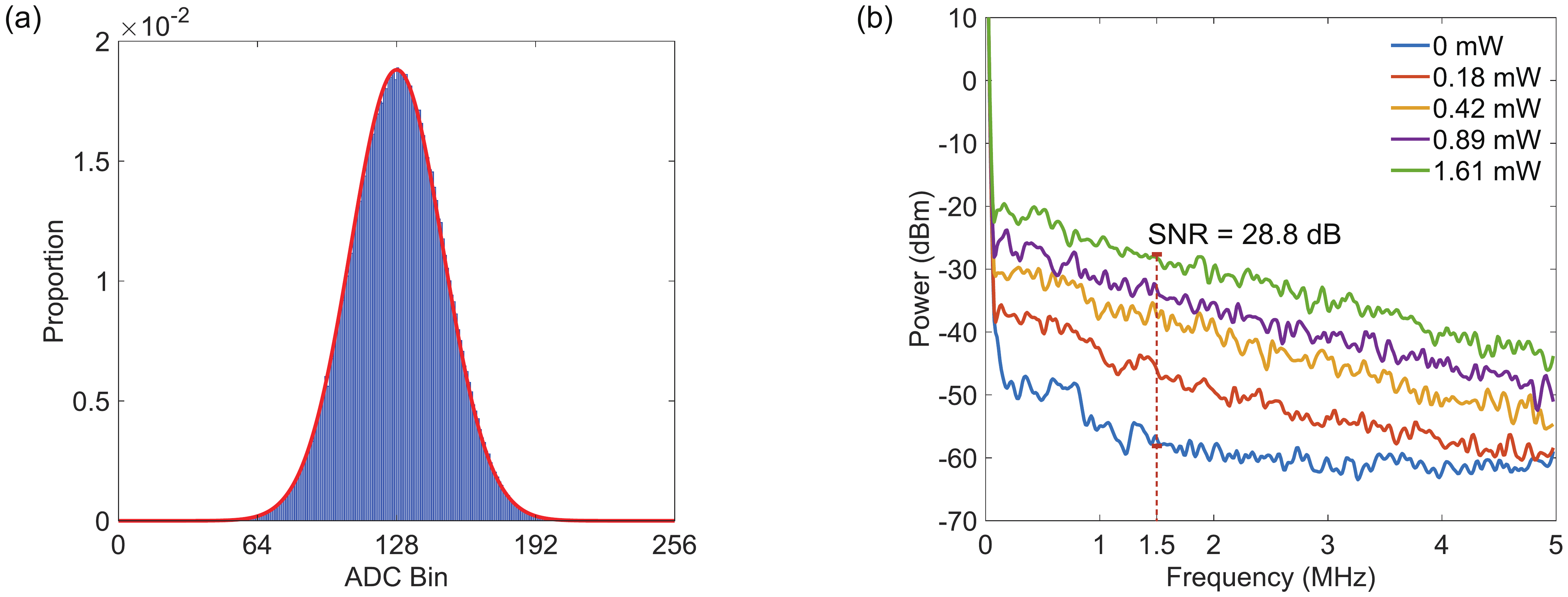}}
    \caption{(a) Histogram and theoretical Gaussian distribution fit of the vacuum-state quantum noise signal collected by the ADC of the MCU. The raw data file consists of $10^6$ samples. (b) Averaged power spectral density measured at the output of the OPA with different settings of LO equivalent power. The ADC sampling rate is 1.5 MSa/s, and the SNR at a bandwidth of 1.5 MHz is 28.8 dB. SNR: signal-to-noise ratio.}
    \label{figure4}
  \end{figure*} 
  
  To verify the functionality of the chip and evaluate its quantum randomness, characterization of the QRNG chip is essential. 
  The evaluation of the raw data requires characterization of the amplitude distribution of the quantum noise signal.
  In the vacuum-state scheme, the amplitude and phase quadratures of the vacuum state are represented by a pair of non-commuting operators, which cannot be simultaneously measured with arbitrary precision.
  In the phase space, the vacuum state is represented as a two-dimensional Gaussian distribution centered at the origin.
  The randomness arises from quadrature measurement of the vacuum-state fluctuations, producing raw data that follows a Gaussian distribution.

  A typical histogram of the raw random data captured by the ADC in the MCU under raw data output mode at an LO equivalent power of 1.6 mW, along with the theoretical Gaussian distribution fit curve, is shown in Fig.~\ref{figure4}(a).
  The raw data consist of $10^6$ samples, and the dynamic range of the 8-bit ADC spans from code 0 to code 255 with a 3.3 V full-scale reference voltage.
  The Gaussian distribution of the raw quantum noise aligns with the principle of vacuum-state fluctuations \cite{gabriel2010generator}.

  The LO equivalent power is defined as the product of the laser power at the LO port and the total light transmission loss from the LD to the homodyne detector.
  A resistor is connected in series at the negative terminal of the LD.
  The LO equivalent power is calculated using the 3.2 dB transmission loss and the operating current of the laser which is obtained by measuring the voltage across the resistor.

  \begin{figure*}[t]
    \centerline{\includegraphics[width=16.5 cm]{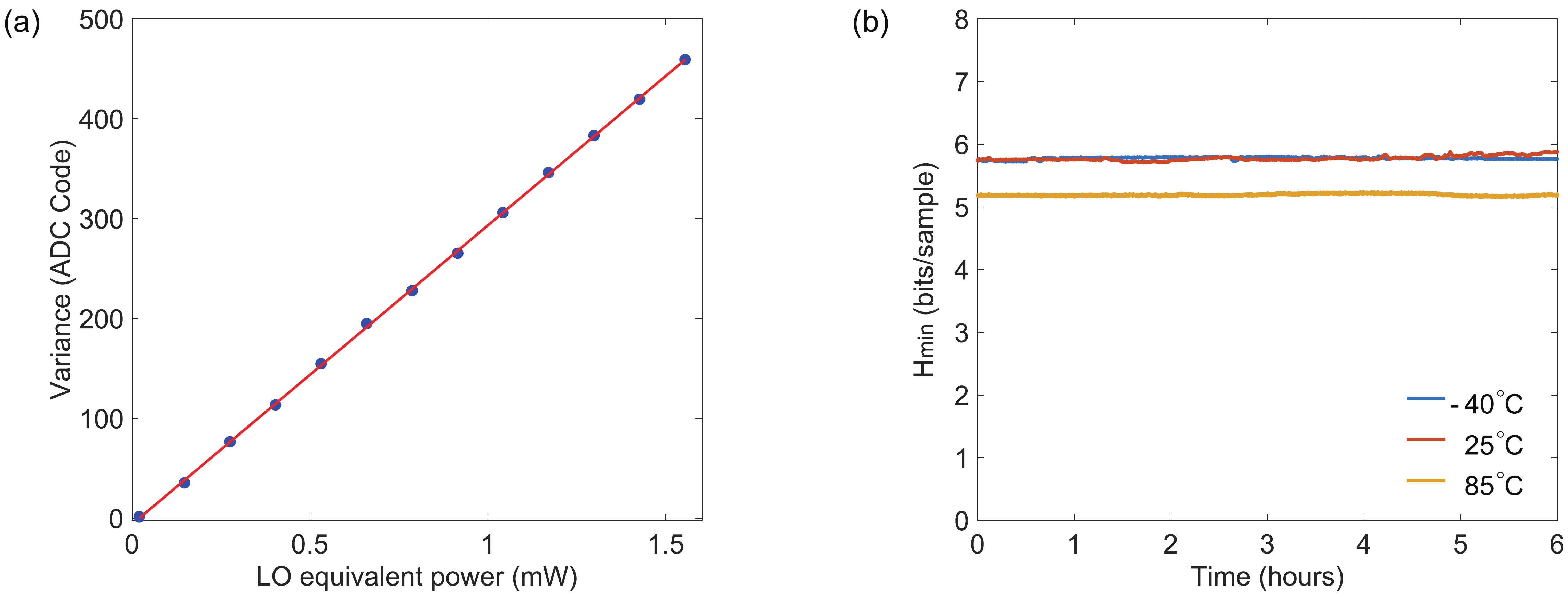}}
    \caption{(a) Variance of the signal amplitude measured by the ADC as a function of LO equivalent power, with the theoretical linear fit shown as the red line. (b) Min-entropy of raw data during 6 hours of continuous operation at -40°C, 25°C, and 85°C.}
    \label{figure5}
  \end{figure*}
  
  To determine the appropriate sampling rate for the ADC and the LD power operating point, the average power spectral density of the quantum noise at various LO equivalent powers is measured using a spectrum analyser.
  The smoothed power spectrum curves, shown in Fig.~\ref{figure4}(b), illustrate the frequency response of vacuum-state fluctuations.
  As the LO equivalent power increases, the quantum noise signal power spectrum proportionally increases.
  The signal-to-noise ratio at an LO equivalent power of 1.61 mW within the bandwidth of 1.5 MHz is 28.8 dB.
  Considering both the frequency response characteristics and the postprocessing limitations of the MCU, the sampling rate of the 8-bit ADC is set to 1.5 MSa/s.

  The contributions of the original noise signal include both quantum shot noise and classical electronic noise; therefore, the quantum randomness of the noise needs to be evaluated.
  Given that both quantum noise and classical noise follow Gaussian distributions, the total variance of the measured signal amplitude is expressed as  $\sigma_{total }^{2}=\sigma_{q}^{2}+\sigma_{c }^{2}$, where $\sigma_{total }^{2}$ represents the total variance measured by the ADC, $\sigma_{q }^{2}$ is the quantum noise contribution, and $\sigma_{c }^{2}$ is the classical noise contribution \cite{shi2016random}.
  The quantum shot noise of vacuum-state fluctuations arises from homodyne detection.
  The classical noise is constant and originates primarily from the thermal noise and dark current noise of the PDs and the amplification circuit.
  By adjusting the LO equivalent power, the amplitude of the original noise signal is regulated to calibrate the randomness of the quantum shot noise.

  Figure~\ref{figure5}(a) shows the experimental results of the variance in the quantum noise for different LO equivalent powers. 
  The ADC code represents the signal amplitude unit, with code 0 and code 255 defining the dynamic range of the ADC.
  The ADC code corresponds to the amplitude of the noise signal, and the full-scale reference voltage of the ADC is 3.3 V.
  For each LO equivalent power setting, the signal amplitude variance is computed from $10^6$ samples. 
  $\sigma_{c }^{2}$ is measured without LO input, and the experimental results show that $\sigma_{c }^{2} = 1.2$ in the ADC code, which is significantly lower than $\sigma_{q }^{2}$ under a typical operating point with an LO equivalent power near 1.6 mW. 
  Compared with the quantum shot noise, the contribution of the classical electronic noise is negligible.
  As shown in Fig.~\ref{figure5}(a), the variance in the quantum shot noise linearly increases with the LO equivalent power, which aligns with the theoretical predictions for vacuum-state fluctuations in the QRNG scheme \cite{bai202118}.

  In the vacuum-state QRNG system, the randomness of the raw data is evaluated using a min-entropy approach.
  The min-entropy $H_{min }(X)$ is calculated as $H_{min }(X)=-\log _{2} P_{max }$, where $P_{max }$ represents the maximum probability, derived from the measured quantum noise variance $\sigma_{q }^{2}$.
  A typical $H_{min }(X)$ is 5.8 bits per sample at an LO equivalent power of 1.6 mW and a room temperature of 25°C.
   
  The reliability and stability of the QRNG chip are critical for practical applications.
  The chip is continuously tested for its operational status over a 6-hour period at -40°C, 25°C, and 85°C.
  The min-entropy evaluation results of the QRNG chip within the industrial temperature range are presented in Fig.~\ref{figure5}(b).
  At -40°C and 25°C, the min-entropy remains stable at approximately 5.8 bits per sample, whereas at 85°C, it stabilizes at approximately 5.2 bits per sample.
  Temperature has a significant effect on the operating conditions of the laser. 
  As the temperature increases, the threshold current of the laser increases, leading to a reduction in the output power. 
  This causes the min-entropy to correspondingly decrease, resulting in a lower min-entropy of the raw data compared with both normal and low-temperature conditions.
   
  The stable long-term operation of QRNG chip at a constant output rate over the industrial temperature range of -40°C to 85°C requires optimization of the randomness extraction ratio.
  Given the serial computing mode of the MCU, a higher extraction ratio of the Toeplitz matrices can improve the utilization of raw data, but it also increases the computational complexity.
  Considering a balance between computational efficiency, temperature tolerance, and overall stability, the randomness extraction ratio is set to 0.5.
  Eight different Toeplitz matrices, each of size 256 × 128, are pre-initialized in the flash memory of MCU and are rotated during postprocessing.
  The seeds of the Toeplitz matrices are periodically refreshed with the final output random numbers.
  At this extraction ratio, the randomness of the raw data must exceed 4 bits per sample within an 8-bit ADC range.

  The randomness threshold is set to 4 bits per sample in the entropy source monitoring function of the MCU and recalibration of the laser operating point is triggered when the temperature fluctuation exceeds 5°C.
  The laser light intensity is adjusted by the DAC of the MCU based on the randomness threshold, PD currents, and temperature sensor readings, and the LD operating point is recalibrated through feedback adjustment.
  The recalibration of the LD ensures that the min-entropy level exceeds the required extraction ratio of 0.5.
  The chip exhibits better min-entropy performance at low temperatures, and it can be optimized for a higher data output rate.
  However, to increase the practicality of the QRNG chip, the data output rate should remain constant across different temperatures.
  The QRNG chip sustains a constant output rate of 5.2 Mbps across the industrial temperature range of -40°C to 85°C, with a power consumption of approximately 200 mW.
  
\begin{figure*}[t]
    \centerline{\includegraphics[width=16.5 cm]{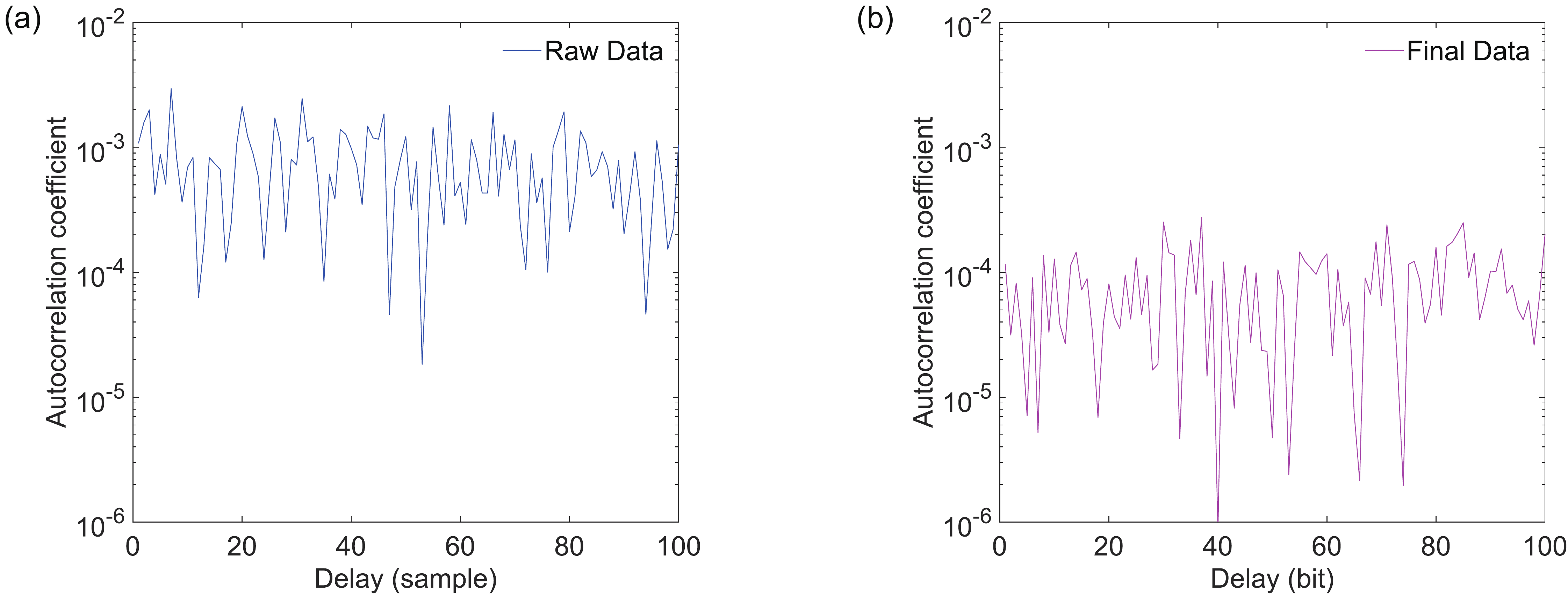}}
    \caption{(a) The autocorrelations of raw random data. The data size of the file is $10^6$ samples. (b) The autocorrelations of the final random data after extraction. The data size of the file is $10^6$ bits.}
    \label{figure6}
  \end{figure*}

  To evaluate the randomness of the final data, an autocorrelation coefficient comparison between the raw data and the extracted random data is conducted.
  The autocorrelation function quantifies the dependence of a signal across different time points.
  As shown in Fig.~\ref{figure6}(a), the autocorrelation coefficient of the raw data remains stable as the sample delay increases. 
  After real-time randomness extraction, the autocorrelation coefficient of the final random data, presented in Fig.~\ref{figure6}(b), is significantly reduced compared with that of the raw data.
  This reduction in the autocorrelation coefficient indicates that the postprocessing successfully enhances the uniformity and unpredictability of the final random sequence.
  The autocorrelation curves exhibit no discernible patterns, confirming statistical independence.
  Additionally, standard NIST statistical tests \cite{NIST} validate the randomness of the final data, with all p-values exceeding 0.01 and all proportions exceeding 0.98.
  The quantum random number data generated at -40°C, 25°C, and 85°C successfully pass all test items.

  \section{Conclusion}
  In summary, we have first reported a vacuum-state QRNG chip that achieves a constant real-time output rate of 5.2 Mbps across the industrial temperature range of -40°C to 85°C, with a size of 16.6 mm × 7.8 mm.
  The vacuum-state entropy source is implemented using a hybrid photonic integration approach via a \ce{SiO2} Y-splitter waveguide.
  Hybrid photonic and electronic components are further integrated into a compact ceramic package using SiP technology.
  Raw data acquisition, postprocessing, and final random number output are managed by the MCU.
  Our work provides a fully functional chip-level solution for vacuum-state QRNGs offering the advantages of cost-effectiveness and low-power consumption, which are suitable for broad deployment in practical applications.

\begin{acknowledgments}

  This work has been supported by the National Key R\&D Program of China (2020YFA0309704), the National Natural Science Foundation of China (12304562, 62275239, 62175227), the Innovation Program for Quantum Science and Technology (2021ZD0300804), the Key R\&D Program of Anhui (202203a13010002), the China Postdoctoral Science Foundation (2023T160614, 2022M713058), the Anhui Provincial Natural Science Foundation (2308085QA27), the USTC Research Funds of the Double First-Class Initiative (YD9990002031).

\end{acknowledgments}

\bibliography{sip_v10}

\end{document}